\documentclass{vgtc}                          

\graphicspath{{figures/}{pictures/}{images/}{./}} 

\newcommand{\kanji}[2][\dimexpr\ht\strutbox+\dp\strutbox\relax]{%
  \raisebox{-\dp\strutbox}{%
    \includegraphics[totalheight=#1]{#2}%
  }%
}

\usepackage{times}                     

\usepackage{tabu}                      
\usepackage{booktabs}                  
\usepackage{lipsum}                    
\usepackage{mwe}                       

\usepackage{mathptmx}                  

\onlineid{0}

\vgtccategory{Research}

\vgtcinsertpkg

\title{Kokatsuji: A Visualization Approach for Typographic Forensics of\\Early Japanese Movable Type}

\author{Ignacio Pérez-Messina\thanks{e-mail: ignacio.messina@tuwien.ac.at}\\ %
        \scriptsize TU Wien, Austria %
\and Asanobu Kitamoto\thanks{e-mail: kitamoto@nii.ac.jp}\\ %
     \scriptsize National Institute of Informatics, Japan %
     }

\teaser{
  \centering
  \includegraphics[width=\linewidth]{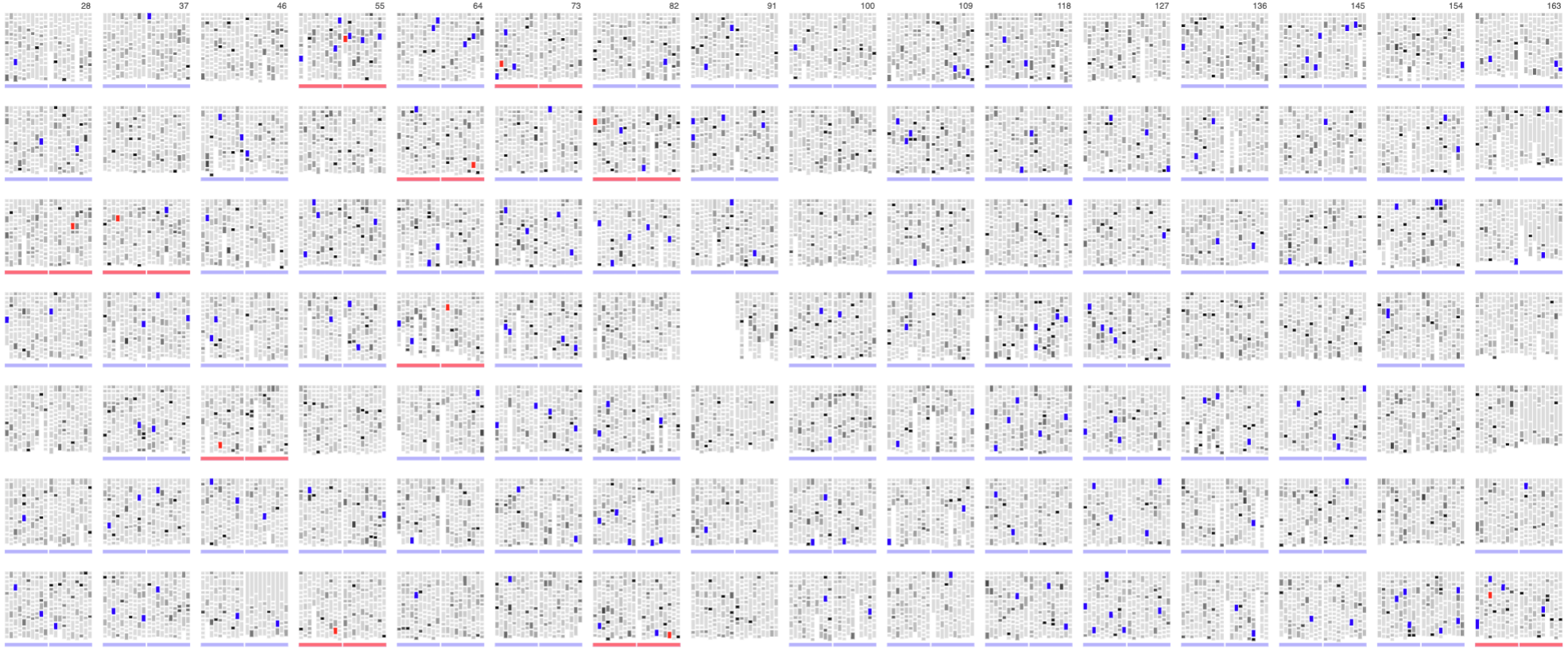}
  \caption{Detail of the Overview of the CODH Sagabon Tsurezuregusa dataset enhanced with automatically identified printing blocks \cite{li2024kokatsuji}. The spreads are arranged in top-to-bottom, left-to-right order. Each is rendered as a composite glyph of its segments, with a white-to-black scale indicating the number of uses of each identified block. Highlighted in red are occurrences of cluster \#14 of character \kanji{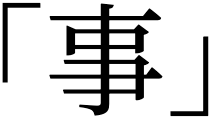}~(\textit{koto}), all tightly localized between spreads 30 and 88 (with a single outlier on spread 169), while in blue are instances of the other block variations sharing the same reading, which are found throughout the book.}
  \label{fig:teaser}
}

\abstract{
    We present a visualization system designed to support typographic forensics in the study of Kokatsuji, the short-lived tradition of Japanese movable wooden type printing. Building on recent advances in machine learning for block identification, our system provides expert users with an interactive tool for exploring, validating hypothesis, and integrating expert knowledge into model-generated results about the production process of early printed books. The system is structured around an ontology of four conceptual objects (spreads, segments, blocks, and characters) each corresponding to a dedicated view in the system. These coordinated views enable scholars to navigate between material evidence and computational abstractions, supporting close, near-by, and distant reading practices. Preliminary results from expert use of the system demonstrate its ability to reveal errors in segmentation, inconsistencies in clustering, and previously inaccessible patterns of block reuse. 
} 


\begin{document}


\firstsection{Introduction}

\maketitle


In the mid-15th century, Gutenberg introduced movable metal type to Europe with the intent to replicate the visual elegance of handwritten manuscripts at scale. His invention ushered in a revolution in the West, making books cheaper and more widely available than ever before, so contributing enormously to the historical passage from the Middle Ages into Modernity. When this technology arrived in Japan more than a century later, it encountered a profoundly different linguistic and cultural landscape. The Japanese writing system, with its thousands of kanji and flowing cursive script (kuzushiji), did not lend itself easily to typographic reduction. Characters often connected fluidly across lines, creating countless visual variations. As a result, Japan’s earliest experiments with movable type, known as kokatsuji, required a level of typographic complexity that rivaled or exceeded that of traditional woodblock printing. The books produced during this brief period were celebrated for their beauty but were expensive and difficult to produce. Unlike in Europe, where printing helped dismantle the gatekeeping of knowledge, kokatsuji did not become a tool for mass communication. It was ultimately abandoned after a few decades, leaving behind a small number of exquisitely printed volumes and a host of unresolved questions for bibliographers.

Among the printed artifacts left by this short-lived experiment, some of the most studied are the Sagabon editions of the early 17th-century. One such work is Tsurezuregusa (``Essays in Idleness'' in Keene's canonical translation~\cite{keene1967essays}), a 14th-century collection of philosophical reflections by Yoshida Kenkō. While the text itself has been preserved and widely studied, the material process behind these kokatsuji editions remains elusive. Scholars of literature and bibliography have hypothesized that different sets of movable type blocks were used to print distinct sections of these books. Certain characters appear with subtle variations (e.g., slightly shifted shapes, uneven inking, or changes in stroke dynamics) suggesting that multiple blocks were carved for the same characters. These differences may point to shared block sets, replacements, or typographic substitutions introduced during the production process. Reconstructing these hidden relationships is a central question in current research of kokatsuji, one that has traditionally required time-consuming manual comparison of printed characters across hundreds of pages.

In this work, we present an interactive visualization system for the forensic investigation of Kokatsuji typography. Building on prior results~\cite{li2024kokatsuji}, our approach enables researchers to visually explore and interact with thousands of segmented character images from Tsurezuregusa, revealing patterns of reuse, anomalies, and typographic variance. Through a set of coordinated views and intuitive interactions, users can navigate an overview of the whole dataset, examine the spatial distribution of character clusters, look at the source material, and so validate and generate hypotheses about the logic of movable type use.

This system introduces a novel visualization to the emerging field of computational typographic forensics. It bridges algorithmic analysis and human interpretation, offering scholars an interactive means to engage with complex, uncertain humanities data. Two main questions drive the design of this system: (1) What is the set of blocks from which this book was printed? and (2) did these block themselves belong to different sets which were used to print different sections of a book? To our knowledge, this is the first work to apply visualization methods to the study of kokatsuji and opens new directions for humanistic inquiry in Japanese book history.







\begin{figure}[t]
  \centering
  \includegraphics[width=\linewidth]{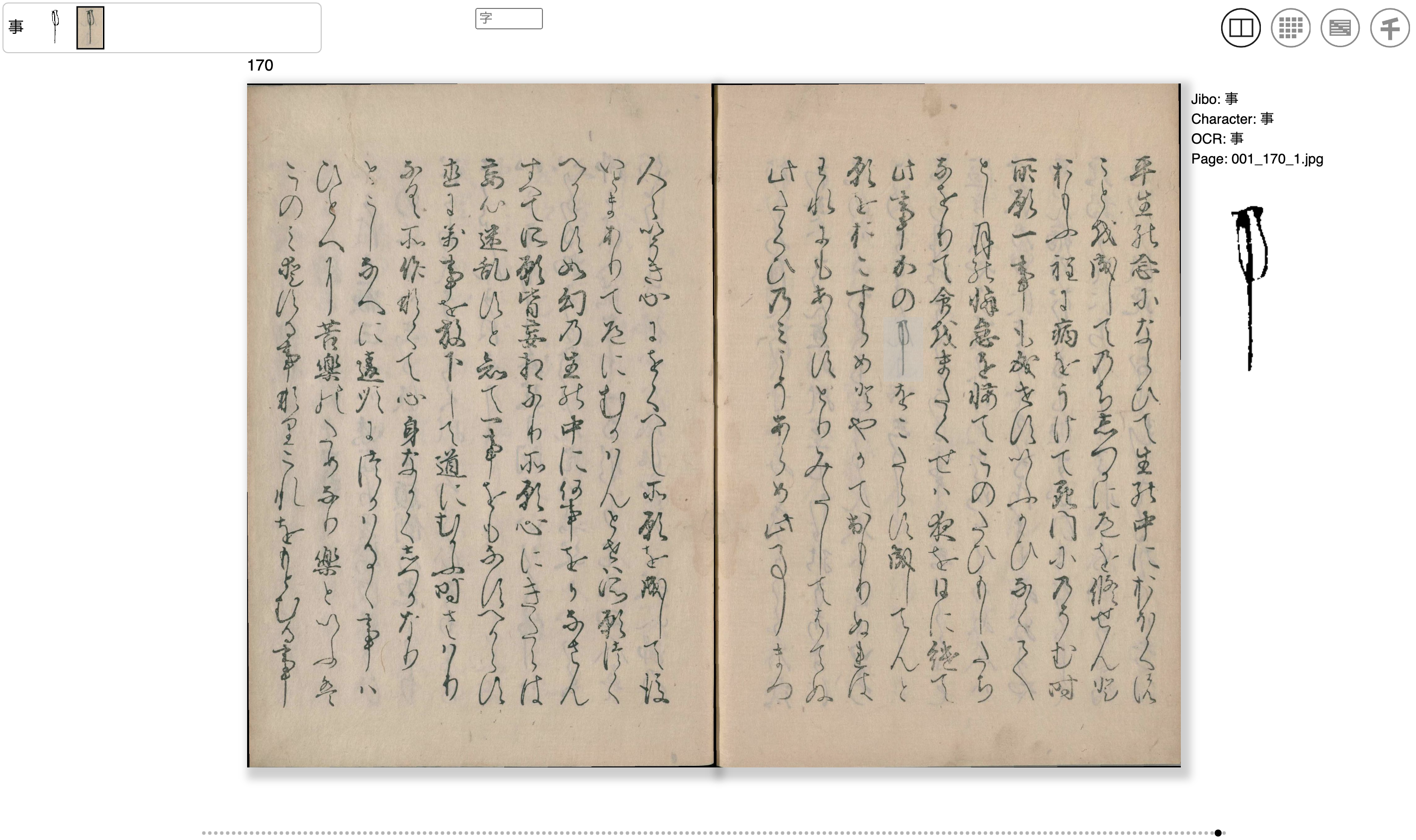}
  \caption{The Source View displaying high-resolution scans of the printed book with overlaid segmentation results. Highlighted is one particular segment from the character \kanji{koto.png}~(\textit{koto}). Users can inspect segment metadata on hover and navigate across spreads to conduct close reading within the original material context.}
  \label{fig:source-view}
\end{figure}

\section{Dataset}
Tsurezuregusa, attributed to the 14th-century monk Yoshida Kenkō, is one of the foundational texts of Japanese classical literature. Composed of 243 short, loosely connected aphorisms reflecting on aesthetics, impermanence, and the habits of the time, Tsurezuregusa became a canonical work in Japanese education and cultural memory. It has been copied, illustrated, and printed in countless versions over centuries. Among these, the Sagabon edition (so named for its association with the Kyoto-based Saga domain) is considered a pinnacle of early Japanese typography. Printed between c. 1600 using Kokatsuji (old movable wooden type), the Sagabon editions of classical texts are admired for their visual beauty, refined layout, and the use of cursive typefaces that mimic brush writing. Unlike wooden block printing, Kokatsuji required the individual composition of characters using a large set of movable type blocks, making its typographic process far more complex and elusive than Western letterpress editions.

To study Kokatsuji from typographic research point of view, the Center for Open Data in the Humanities (CODH) released the Kokatsuji Dataset\footnote{\url{https://codh.rois.ac.jp/omt/dataset/}} for the Sagabon Tsurezuregusa with the CC BY open data license. Based on the complete description of the pipeline for creating the dataset \cite{jm23a}. The following offers a summary of the dataset.

First, segments were extracted from high-resolution scans of the pages of the Tsurezuregusa Sagabon edition. The segmentation process was based on the assumption that each segment has a height that is a multiple of a unit length and a uniform width. While each segment may contain one or more characters, individual characters within a segment can vary in height, resulting in a different number of characters per segment even when the segment height is the same. A crucial feature used in segmentation was the presence of cursive script, as connected characters are assumed to belong to the same printing block. Segments were therefore validated by checking whether the unit-length segmentation could be applied without splitting connected characters. As a result, each page image was represented as a set of horizontally aligned lines with the same width, and each line as a vertically aligned set of segments with varying heights. A total of 36,869 segments were extracted from the Sagabon Tsurezuregusa.

Second, each segment was annotated with its corresponding Unicode character. Additionally, the\textit{jibo} field was included for kana characters to represent their original mother kanji. This information allowed the system to distinguish between characters that are mapped to the same Unicode value but originate from distinct historical forms.

Based on this annotated dataset, a derivative dataset was created to support the study of character shapes for the identification of distinct printing blocks. Clustering was applied to segments representing the same character, using visual similarity to approximate block identity~\cite{li2024kokatsuji}. The clustering pipeline combined structural similarity metrics (SSIM), deep learning-based contrastive features, and community detection methods to group visually similar segments. These results enabled statistical analyses such as the identification of block reuse across pages and potentially across books. However, due to the lack of direct physical evidence, the reconstructed blocks remain hypothetical and require further evaluation by domain experts.

The dataset provides a foundation for further investigation by combining source images, segmentation results, character identification, and block reconstruction. While much of the work in typographic forensics has focused on the computational models for creating these datasets (for example, in medieval manuscripts~\cite{boudraa2024integrating,lastilla2022self,stutzmann2016clustering} and historical books~\cite{im2022deep,lacasta2022tracing,thomas2024capturing}) the present work extends these results by adding a new layer of interactive visualization tools that enable experts to assess and refine these computational interpretations critically.



\section{Visualization Approach}

From a visual analytics perspective, we follow the data–users–tasks triangle~\cite{miksch2014matter}. \textbf{Data}: the CODH Kokatsuji dataset augmented with automatic segmentation and clustering that hypothesize block identity. \textbf{Users}: literary scholars reconstructing Kokatsuji production. \textbf{Tasks}: reconstruct block identity and analyze reuse patterns. To address these requirements, the interface is designed to support the four search tasks (lookup, browse, locate, explore)~\cite{brehmer2013multi} across the ontology of characters, segments, and blocks. Our expanding multi-level selection schema (where selections at any level propagate to adjacent levels) provides coordinated interaction and sustains coherence while moving between detail and overview.

However, applying standard visualization paradigms requires reconsideration in the digital humanities context. While much of this field has been structured around the complementary opposition of close and distant reading~\cite{janicke2015close}, this framing has traditionally assumed a literal, textual conception of reading. In our project, despite belonging to the domain of textual cultural heritage, the object of investigation is not the text itself, but the typographic traces left by movable type. Nevertheless, reading still takes place, though it operates at the indexical level of the sign, much like interpreting a photograph. Here, meaning is constructed not by deciphering linguistic content, but by examining visual evidence that connects printed traces to theorized physical objects: the movable type blocks. In this sense, our system functions as a tool for typographic forensics.

Our interface design reflects this progression from material evidence to abstraction. It begins by grounding the user in the scans of the printed book, facilitating close reading of the impressions themselves: the shapes, textures, and imperfections that constitute the physical remnants of the printing process. Additional data layers derived from computational analysis introduce interpretive hypotheses: segmentation delineates the estimated boundaries of individual block impressions, while clustering proposes which impressions may originate from the same physical block. Though speculative, these abstractions enable the construction of overview visualizations that support distant reading by surfacing large-scale reuse patterns and structural tendencies.

These layers of computationally generated data are not neutral. They encode specific algorithmic interpretations of the material, and their validity remains subject to scholarly scrutiny. The system is thus not merely a neutral tool but a space for critical engagement, where expert knowledge and algorithmic outputs intersect. Within this spectrum, the notion of near-by reading~\cite{mcnutt2020supporting} provides a productive conceptual frame. Our expert users, already familiar with the material and its historical context, leverage the system's abstract visualizations to navigate between large-scale patterns and fine-grained typographic details. The system enables them to move fluidly between levels of evidence, maintaining interpretive agency throughout.

\begin{figure*}[th!]
  \centering
  \includegraphics[width=\linewidth]{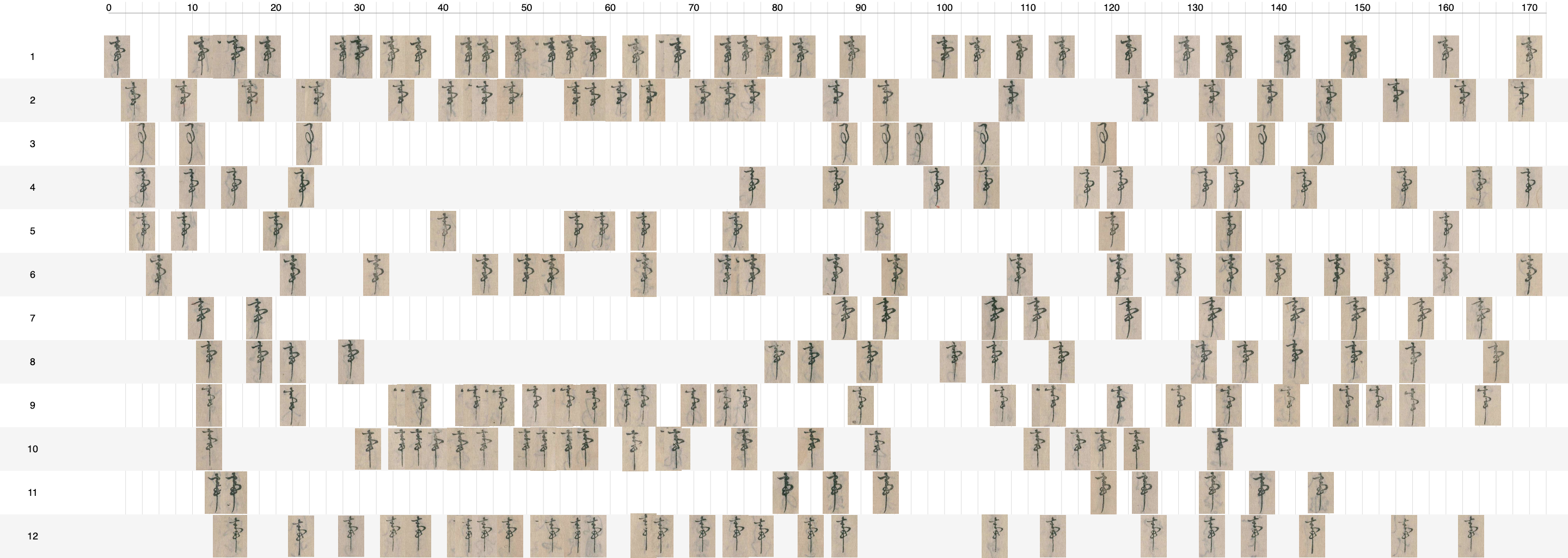}
  \caption{Detail of the Analytical View for the character \kanji{koto.png}~(\textit{koto}) showing the first 12 clusters out of 34. Each row corresponds to a cluster of segments (a block variant for the character) and their appearances across the book. A clear partition in usage pattern can be observed between the left and right sides of the plot, having its origin exactly at spread 95 where the second volume of the book starts. This view also allows for drag-and-drop editing of the block-segments.}
  \label{fig:analytical-view}
\end{figure*}

\subsection{Ontology}

The design of our system is grounded in an ontology that reflects both the material structure of the Kokatsuji printing process and the interpretive layers introduced by computational analysis. Four main object types constitute this ontology: spreads, segments, blocks, and characters. These entities form the conceptual backbone of our visualization and define the relationships that the system makes visible to expert users. Characters, blocks and segments are directly selectable across the different views of the system, and the selection at each of these levels expands upwards and downwards in the ontological hierarchy. 

\noindent\textbf{Spreads.}
A spread refers to the double-page layout commonly used in graphic design. However, in the context of early Japanese bookbinding, its structure is technically different. Traditional Japanese books were printed on one side of the paper, which was then folded to create pages on both sides, leaving the interior of the fold blank. As a result, when the book is open, the two facing pages do not belong to the same physical sheet but to consecutive spreads. This distinction is critical for typographic forensics. The spread represents the largest reconstructable unit of the printing process, corresponding to a specific arrangement of type blocks assembled for a single impression. Within a spread, the same physical block cannot logically appear twice, while co-occurring blocks provide evidence for being part of the same subset of the printer's inventory.

\noindent\textbf{Segments.}
A segment refers to an extracted region of the digitized page image that is interpreted to contain the imprint of a single type block. Segments are not direct representations of the physical trace itself, but rather the result of an automated segmentation process applied to the scanned book images. This process, while necessary for computational analysis, introduces an interpretive layer that requires critical scrutiny. Each segment is assumed to correspond to the imprint of a single block and contains one or more characters chained together. Segments thus constitute the primary units of analysis for block identification, but the validity of each one a as faithful representation of the underlying printing process remains contingent on expert evaluation.

\noindent\textbf{Blocks.}
A block refers to a type block crafted for kokatsuji printing, analogous to the movable type used in Gutenberg-style letterpress printing. Unlike their European counterparts, kokatsuji blocks were made of wood, larger in size, and designed for vertical text layout. Almost no physical examples of these blocks have survived (particularly, of the Sagabon editions), making them inherently hypothetical entities. They remain the central object of inquiry for typographic forensics in this context, their existence inferred only through the recurring patterns observed in segments.

\noindent\textbf{Characters.}
A character or characters refer to the written linguistic units that is the content of a segment or block. Characters allow for the grouping of blocks into ``variations'' of a same linguistic unit (i.e., allographs). In the case of Western letterpress, the set of characters would mainly be constituted by upper and lower case single characters with some chains of characters that often appear together. In kokatsuji the case is much more complex, as almost 4000 different character sequences are present in the dataset.

\begin{figure}[th]
  \centering
  \includegraphics[width=\linewidth]{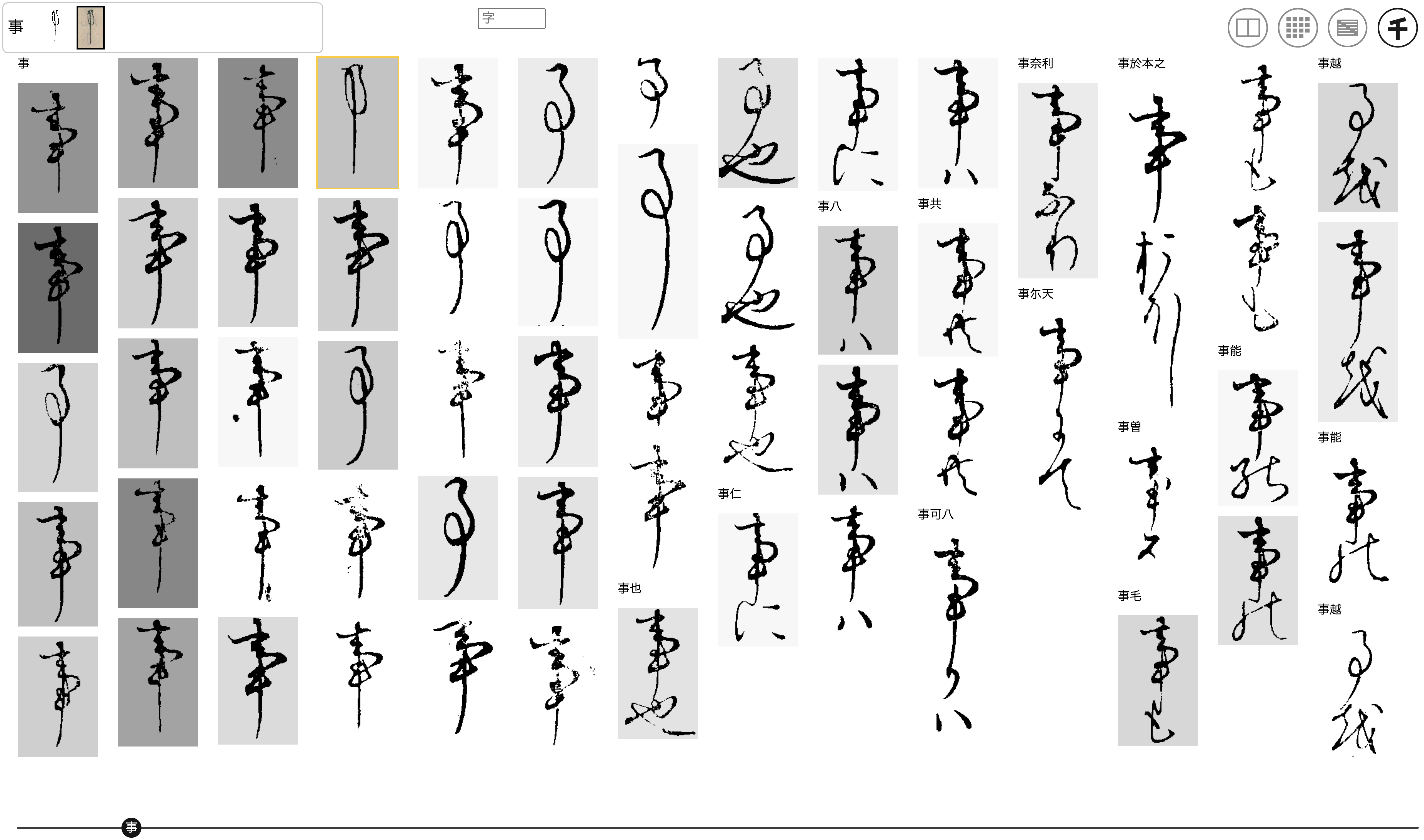}
  \caption{The Collection View displaying inferred blocks for character \kanji{koto.png}~(\textit{koto}) and subsequent characters. The background color of the block encodes their number of appearances, from white (1) to darker (many). The block thumbnails are thresholded segment images to create block images abstracted from their original page context.}
  \label{fig:type-case}
\end{figure}

\subsection{Views}

Our visualization approach consists of four main views, each corresponding to a distinct function relating to each of our ontological entities. The Source View focuses on the material layout of the book, providing access to the high-resolution scans, enabling close reading. The Overview displays all segmented impressions across the book, offering a comprehensive space for selecting and exploring segments and their connections. The Analytical View enables exploration and correction of block variants within individual characters, combining pattern inspection with direct editing of clustering results. Finally, the Collection View presents the inferred blocks as first-class entities, abstracted from their page context, allowing for the evaluation of block-variants and cluster coherence.

\noindent\textbf{Source View.} The Source View begins with a direct simulation of the book. High-resolution scans of the full pages are arranged side by side as if they were physically bound together, with navigation facilitated by a visual slider that allows users to ``turn'' pages. Over this material foundation, a layer of model results is superimposed, revealing the automated segmentation of character impressions, which can be inspected on hover. Originally conceived as an optional component, the Source View has ultimately become central to enabling close reading practices within the system. From any other view, users can right-click on a segment to access its corresponding spread. This provides not only spatial context for interpreting the segment but also access to segment metadata on hover. The Source View also offers an alternative navigation pathway: selecting a segment within the spread returns the user to the Overview, with that segment pre-selected for further inspection (Fig.~\ref{fig:source-view}).

\noindent\textbf{Overview.} The Overview allows for a complete view of the dataset (Fig.~\ref{fig:teaser}). It displays each spread of the book as a composite glyph: the glyph is made of several independent glyphs corresponding to the identified segments as a simple rectangle that conserves the aspect ratio of the segment. Composite glyphs where previously explored by Pérez-Messina et al. for archival aerial images~\cite{perez2023guided} and writing processes~\cite{perez2018organic}. This abstraction allows not only to get a full view of the dataset in one screen, but to encode different layers of derived information in the coloring of the segment glyphs. However, the most important feature of this view is to make visible the underlying network of blocks. This relational space is not directly visualized but revealed through interactive hovering and selection of segments inside the spread glyphs. Following a similar concept to that of monads~\cite{dork2014monadic}, this dense layer of interconnections is only visible from the partial perspective of its particular elements. If the glyphs were freed from their fixed positions in the composite glyph following the material's original form, this arrangement could give rise to a fluid visualization where all information is encoded in the layout of the elements (see, for example, Coins by Gortana et al.~\cite{gortana2018off}), however, this design possibility remains a future area of exploration.

\noindent\textbf{Analytical View.} The Analytical View is focused on all the block variations for a single character structured as a block instance distribution chart (Fig.~\ref{fig:analytical-view}). It displays the different identified block variants as horizontal rows, and their appearances across the book along the X-axis. Beyond providing an overview of block reuse and distribution patterns for individual characters, this view integrates cluster editing functionality. Segments can be interactively dragged from one cluster to another, clusters can be merged, and these changes are immediately reflected across the system's other views. To support informed decisions during this process, hovering over a segment thumbnail reveals a full-resolution image of the corresponding segment, enabling close inspection of subtle typographic details. The modified state of the dataset can be exported as a new JSON file, facilitating iterative expert curation and reproducible analysis across sessions. We believe this data editing capabilities are important not only for the integration of expert knowledge into the algorithmic results but for the increase of trust from scholars in the visualization itself by ``by-passing'' ``black-boxing'' and ``the lure of objectivity'' (see van den Berg et al.'s discussion on Rieder and Röhle's challenges in the digital humanities~\cite{van2018philosophical,rieder2012digital}).

\noindent\textbf{Collection View.} The Collection View presents the inferred blocks as first-class entities, arranged in an orderly, searchable layout reminiscent of how type blocks would be organized for printing as in what was originally called a \textit{type case} (Fig.~\ref{fig:type-case}). Blocks are grouped by character and displayed alongside their most relevant derived feature: their number of uses. This is visually encoded through background color, ranging from white (single-use blocks) to progressively darker tones, following expert feedback that ink traces tend to darken with repeated use. A filter allows users to restrict the view to blocks within specific reuse ranges, supporting targeted exploration. To visually differentiate blocks from individual segments, the system selects a representative segment from each cluster and applies a lightness threshold to extract only the ink imprint, abstracting it from its original page and reaffirming its status as a hypothetical, reconstructed entity.

A demo version of the system will be made publicly available\footnote{\url{http://kokatsuji.baltazarperez.com/demo/}}.

\section{Preliminary Results}
A preliminary visual inspection of the overview interface revealed immediate issues in the underlying segmentation data. In particular, several pages exhibited character segments with abnormally large bounding boxes suggesting clear errors in the automatic segmentation process (see, for example, spreads 52, 164, 165 and 167 in Fig.~\ref{fig:teaser}). These anomalies, difficult to detect through tabular inspection alone, became conspicuous when the spatial distribution of segments was rendered at scale.

It also became apparent through the Collection View that many segments marked as single-use blocks likely belong to larger clusters, but variations in ink shading or print quality prevented the computer vision and clustering pipeline from grouping them together. This highlights the importance of providing a user-friendly cluster editor, allowing experts to correct such mistakes. While manually reviewing every cluster remains laborious, experts are typically interested in a specific set of characters relevant to their research, making targeted corrections both feasible and essential. 

On the other hand, applying a custom colormap to highlight same-spread block reuses in the Overview revealed apparent anomalies distributed across the dataset. At first glance, these seemed to contradict one of the fundamental assumptions of typographic forensics: that a physical block cannot appear twice within the same spread. However, close inspection by experts confirmed these cases to be clustering errors. The differences between certain blocks are often extremely subtle, as multiple blocks were carved using nearly identical models. As a result, visual similarity alone occasionally overstates block identity, reinforcing the need for expert validation alongside automated methods. 

The Analytical View, featuring a simple ``timeline'' for different block-variants of a character, shows potential in the unveiling of block usage patterns and generation of new hypotheses. As in the case depicted in Fig.~\ref{fig:analytical-view}, showing the distribution of one of the kanji of most interest for scholars, different patterns are clearly distinguishable for each volume of the book. The interpretation of this finding is still subject to expert consideration.

While the original clustering results by Li and Kitamoto~\cite{li2024kokatsuji} already revealed an expected Zipfian distribution of block reuse (with most blocks used only once or a few times and a small number reused extensively) the interactive selection and highlighting features of our system exposed a more complex structure. By tracing block reuse across spreads, it became apparent that nearly all 171 spreads are interconnected through shared blocks, forming what is effectively an almost complete clique. This observation challenges existing scholarly theories suggesting that distinct partitions of blocks, possibly owned by different parties, were used to print different sections of the book. At the same time, it calls for more fine-grained analysis to determine whether certain spreads exhibit stronger connections than others, potentially revealing subtler patterns of typographic organization or production history.

\section{Discussion}
One of the most immediate outcomes of this project has been the exposure of uncertainty in the original dataset. While our current visualization pipeline does not explicitly encode uncertainty, it has nevertheless made visible the interpretive limitations embedded in the preprocessing stages of the dataset. Errors in segmentation, cluster assignments, and implausible use patterns surfaced through visual exploration, revealing how data-related uncertainties propagate through the computational analysis and into the interface.

In this context, the role of visualization extends beyond simply presenting data. It actively reveals the constructed and uncertain nature of the dataset itself. Yet, our interdisciplinary collaboration with expert literature scholars has also demonstrated that users approach the system with this awareness already in place. The visualized results are not interpreted as objective truth but as another layer of interpretation, distinct from their own and subject to scrutiny. This epistemic stance suggests that introducing explicit uncertainty visualization might not provide a favorable trade-off in this setting. The added complexity of implementing uncertainty representations could outweigh potential gains in trust or informativeness, especially when expert users already engage with the system from a fundamentally critical perspective.

The observations regarding the densely connected spread structure illustrate how visual analysis can surface contradictions within existing scholarly theories. Whether these contradictions reflect errors in clustering or overlooked complexity in historical production practices remains an open question at this point, reinforcing the need for continued interdisciplinary collaboration.

A more fine-grained investigation into these theories could be pursued by extending the current system along the spread-connection metaphor. Specifically, by defining a co-appearance matrix that captures how often blocks appear together across different spreads, it becomes possible to explore the structure of block reuse at a higher level of abstraction. Applying dimensionality reduction techniques to this matrix would allow the construction of a spatial map of blocks, revealing latent relationships, groupings, or partitions that may remain hidden in the current representation. This would not only introduce a novel mode of inquiry into this digital humanities collaboration but also provide a quantitative framework to critically assess existing hypotheses about block ownership, reuse patterns, and production organization.

Yet another aspect of interest to the experts concerns the use of blocks of different length: different block variants can be longer or shorter, and that the choice of blocks is related to the aesthetic principle of always filling up a column perfectly. Their observation that a certain ``rythm'' in the length of blocks can be identified as a significant variable in the printing process, could be tackled in visualization with a similar method as the one found in literature fingerprinting~\cite{keim2007literature}, where a divergent colorscale is applied to sections of text abstracted as equal-sized squares forming a composite glyph, reminiscent of our encoding for segments in the Overview~\ref{fig:teaser}. In our case, re-encoding the length of the segments as their fill color could show patterns that are impossible to observe with the methods currently used by the experts in this topic.

These observations reinforce a broader view of visualization not merely as a neutral analytical tool, but as a negotiation space: an environment where algorithmic interpretations, expert knowledge, and material evidence intersect. In the context of typographic forensics, where the object of inquiry itself is hypothetical and material traces are fragmentary, such a space becomes essential. Visualization in this project does not provide final answers; instead, it facilitates the process of evaluating competing interpretations, exposing ambiguities, and guiding expert attention to areas of uncertainty or contradiction. The design process and interdisciplinary collaboration has made these epistemic tensions visible and actionable, supporting the iterative construction of a tool for inquiry rather than implying closure to the questions driving this research program.

While this system was designed for the analysis of a single book, the long-term trajectory of this research necessarily points toward cross-book, corpus-level comparison. Expanding the inquiry beyond individual volumes would allow for a much broader reconstruction of production practices with kokatsuji. The amount of digitized material already available provides a foundation for this next step, but its exploration remains a challenge for future work.

\section{Conclusion}

This project presents the first application of visualization to the domain of kokatsuji, introducing an interactive system to support typographic forensics in early Japanese movable type books. By combining computational results with visual exploration, the system enables scholars to inspect, question, and refine hypotheses about block reuse and production practices. Through integrated close reading capabilities, editable cluster structures, and coordinated views, the system provides a practical tool for critical inquiry into the material and historical logic of kokatsuji. While currently applied to a single book, it establishes a foundation for future corpus-level analysis and broader investigations into the early application of the Gutenberg press in Japan.

\acknowledgments{
The authors wish to thank Prof. Shunsuke Kigoshi of the National Institute of Japanese Literature, and Prof. Dan Koakimoto of Hosei University for valuable suggestions as domain experts. This work is supported by the International Internship Program from the National Institute of Informatics, and by Vienna Science and Technology Fund (WWTF) under grant [10.47379/ICT19047].}

\bibliographystyle{abbrv-doi}

\bibliography{template}
\end{document}